\documentclass[12pt]{article}
\usepackage{graphicx}
\usepackage{amsmath}
\usepackage{cite}
\newcommand{\ba}{\begin{array}}
\newcommand{\ea}{\end{array}}
\newcommand{\be}{\begin{equation}}
\newcommand{\ee}{\end{equation}}
\makeatletter 

\begin{document}
\baselineskip=24pt
\thispagestyle{empty}
\topskip=0.5cm
\begin{flushright}
\begin{tabular}{c}
2nd February 2006
\end{tabular}
\end{flushright}

\vspace{1cm}

\begin{center}

{\Large\bf Neutrino masses and mixings in a Minimal $S_{3}$-invariant
  Extension of the Standard Model.}
\vspace{1cm}
\linebreak
O. F\'elix, A. Mondrag\'on, M. Mondrag\'on and E. Peinado\footnote{
olga$_{-}$flix@yahoo.com.mx\\
\hspace*{0.5cm}mondra@fisica.unam.mx\\
\hspace*{0.5cm}myriam@fisica.unam.mx\\
\hspace*{0.5cm}eduardo@fisica.unam.mx} 


\vspace{0.2in}

{\sl Instituto de    F\'{\i}sica, UNAM,
    Apdo. Postal 20-364,  01000 M{\'e}xico D.F., \ M{\'e}xico.}
\end{center}

\vspace{0.7cm}    

\begin{abstract}\noindent
  The mass matrices of the charged leptons and neutrinos, that had
  been derived in the framework of a Minimal $S_{3}$-invariant
  Extension of the Standard Model (Kubo J, Mondrag\'on A, Mondrag\'on M,
  Rodr\'{\i}guez-J\'auregui E. {\it Prog. Theor. Phys.}  {\bf 109}, 795,
  (2003)), are here reparametrized in terms of their eigenvalues. The
  neutrino mixing matrix, $V_{PMNS}$, is then computed and exact,
  explicit analytical expressions for the neutrino mixing angles as
  functions of the masses of the neutrinos and charged leptons are
  obtained. The reactor, $\theta_{13}$, and the atmosferic, $\theta_{23}$,
  mixing angles are found to be functions only of the masses of the
  charged leptons. The numerical values of $\theta_{13}^{th}$ and
  $\theta_{23}^{th}$ computed from our theoretical expressions are found
  to be in excellent agreement with the latest experimental
  determinations. The solar mixing angle, $\theta_{12}^{th}$, is found to
  be a function of both, the charged lepton and neutrino masses, as
  well as of a Majorana phase $\phi_{\nu}$. A comparison of our
  theoretical expression for the solar angle $\theta_{12}^{th}$ with the
  latest experimental value $\theta_{12}^{exp}\approx 34^{\circ}$ allowed us to
  fix the scale and origin of the neutrino mass spectrum and obtain
  the mass values $|m_{\nu_{2}}|=0.0507eV$, $|m_{\nu_{1}}|=0.0499eV$ and
  $|m_{\nu_{3}}|=0.0193eV$, in very good agreement with the observations
  of neutrino oscillations, the bounds extracted from neutrinoless
  double beta decay and the precision cosmological measurements of the
  CMB.


\end{abstract}

\newpage
\baselineskip= 24pt
\section{Introduction}
The discovery of neutrino masses and mixings marked a turning point in
our understanding of nature and brought neutrino physics to the focus
of attention of the particle, nuclear and astrophysics
communities~\cite{jung}. Recent neutrino oscillation observations and
experiments have allowed the determination of the differences of the
neutrino masses squared and the flavour mixing angles in the leptonic
sector. The solar~\cite{altmann,smy,ahmad,aharmim},
atmospheric~\cite{fukuda,Ashie:2005ik} and
reactor~\cite{bemporad,Araki:2004mb} experiments produced the
following results:
\begin{eqnarray}
7.1\times 10^{-5}(eV)^2 \leq \Delta^2m_{12} \leq 8.9\times 10^{-5}(eV)^2,\\
\nonumber\\
0.24 \leq sin^2\theta_{12} \leq 0.40,\\
\nonumber\\
1.4\times 10^{-3}(eV)^2 \leq \Delta^2m_{13} \leq 3.3\times 10^{-3}(eV)^2,\\
\nonumber\\
0.34 \leq sin^{2}\theta_{23} \leq 0.68,
\end{eqnarray}
at $90\%$ confidence level~\cite{Maltoni:2004ei,schwetz}.
The CHOOZ experiment~\cite{chooz} determined an upper bound for the flavour
mixing angle between the first and the third generation:
\be
sin^{2} \theta_{13} \leq 0.046.
\ee 
Neutrino oscillation data are insensitive to the absolute value of neutrino
masses and also to the fundamental issue of whether neutrinos are
Dirac or Majorana particles. Hence, the importance of the
upper bounds on neutrino masses provided by the searches that probe
the neutrino mass values at  rest: beta decay experiments~\cite{eitel},
neutrinoless double beta decay~\cite{eliot} and precision
cosmology~\cite{elgaroy}.

On the theoretical side, the discovery of neutrino masses and mixings
has also brought about important changes. 
In the Standard Model, the Higgs and Yukawa sectors, which are
responsible for the generation of the masses of quarks and charged
leptons, do not give mass to the neutrinos. Furthermore, the Yukawa
sector of the Standard Model already has too many parameters
whose values can only be determined from experiment. These two facts,
taken together, point to the necessity and convenience of eliminating
parameters and systematizing the observed hierarchies of
masses and mixings, as well as the presence or absence of CP violating
phases, by means of a flavour or family symmetry under which the
families transform in a non-trivial fashion. Such a flavour symmetry
might be a continuous group or, more economically, a finite group.

In a recent paper, we argued that such a flavour symmetry, unbroken at
the Fermi scale, is the permutational symmetry of three objects,
$S_{3}$, and introduced a Minimal $S_{3}$-invariant Extension of the
Standard Model~\cite{kubo1}. In this model, we imposed $S_{3}$ as a
fundamental symmetry in the matter sector. This assumption led us
necessarily to extend the concept of flavour and generations to the
Higgs sector. Hence, going to the irreducible representations of
$S_{3}$, we added to the Higgs $SU(2)_{L}$ doublet in the
$S_{3}$-singlet representation two more Higgs $SU(2)_{L}$
doublets, which can only belong to the two components of the
$S_{3}$-doublet representation. In this way, all the matter fields in
the Minimal $S_{3}$-invariant Extension of the Standard Model - Higgs,
quark and lepton fields, including the right handed neutrino fields-
belong to the three dimensional representation ${\bf 1}\oplus{\bf 2}$
of the permutational group $S_{3}$. The leptonic sector of the model
was further constrained by an Abelian $Z_{2}$ symmetry.

The group $S_{3}$~\cite{Fritzsc1,pakvasa1,Fritzsc2,harari,Fritzsc3,yamanaka,kaus,Fritzsch4,Harrison} and the product
groups $S_{3}\times S_{3}$~\cite{Harrison,mondragon1,mondragon2,xing} and
$S_{3}\times S_{3}\times S_{3}$~\cite{hall,hall2} have been
considered by many authors to explain successfully the hierarchical
structure of quark masses and mixings in the Standard Model. However,
in these works, the $S_{3}$, $S_{3}\times S_{3}$ and $S_{3}\times
S_{3}\times S_{3}$ symmetries are explicitly broken at the Fermi scale
to give mass to the lighter quarks and charged leptons, neutrinos are
left massless. Some other interesting models based on the $S_{3}$,
$S_{4}$ and $A_{4}$ flavour symmetry groups, unbroken at the Fermi
scale, have also been proposed~\cite{koide,ma,ma2,babu,chen,grimus-la}, but in
those models, equality of the number of fields and the irreducible
representations is not obtained.

In this paper, we derive exact, explicit, analytic expressions for
the elements of the leptonic mixing matrix, $V_{PMNS}$, as functions
of the masses of the charged leptons and the neutrinos. By comparison
with the latest experimental data on neutrino mixings, we obtain
numerical values for the neutrino masses in good agreement with the
experimental bounds extracted from the precision observation of the
cosmic microwave background (CMB) and the neutrinoless double beta decay.
\section{The Minimal $S_{3}$-invariant Extension of the Standard
  Model}
In the Standard Model analogous fermions in different generations have
completely identical couplings to all gauge bosons of the strong, weak
and electromagnetic interactions. Prior to the introduction of the
Higgs boson and mass terms, the Lagrangian is chiral and invariant
with respect to permutations of the left and right fermionic fields.

The six possible permutations of three objects $(f_{1},f_{2},f_{3})$
are elements of the permutational group $S_{3}$. This is the discrete,
non-Abelian group with the smallest number of elements. The
three-dimensional real representation is not an irreducible
representation of $S_{3}$. It can be decomposed into the direct
sum of a doublet $f_{D}$ and a singlet $f_{s}$, where
\be
\begin{array}{l}
f_{s}=\frac{1}{\sqrt{3}}(f_{1}+f_{2}+f_{3}),\\
\\
f_{D}^{T}=\left(\frac{1}{\sqrt{2}}(f_{1}-f_{2}),\frac{1}{\sqrt{6}}(f_{1}+f_{2}-2f_{3})\right).
\end{array}
\ee
The direct product of two doublets ${\bf p_{D}}^{T} =(p_{D1},p_{D2})$
and ${\bf q_{D}}^{T}=(q_{D1},q_{D2})$ may be decomposed into the direct
sum of two singlets ${\bf r_{s}}$ and ${\bf r_{s'}}$, and one doublet
${\bf r_{D}}^{T}$ where
\be
\begin{array}{lr}
{\bf r_{s}} = p_{D1} q_{D1} + p_{D2}q_{D2}, & {\bf r_{s'}} =
p_{D1}q_{D2} - p_{D2}q_{D1},
\end{array}
\ee
\be
{\bf r_{D}}^{T}= (r_{D1},r_{D2})=(p_{D1} q_{D2} + p_{D2}q_{D1},p_{D1} q_{D1} - p_{D2}q_{D2}).
\ee
The antisymmetric singlet ${\bf r_{s'}}$ is not invariant under $S_{3}$.

Since the Standard Model has only one Higgs $SU(2)_{L}$ doublet,
which can only be an $S_{3}$ singlet, it can only give mass to the
quark or charged lepton in the $S_{3}$ singlet representation, one in
each family, without breaking the $S_{3}$ symmetry.

Hence, in order to impose $S_{3}$ as a fundamental symmetry, unbroken
at the Fermi scale, we are led to extend the Higgs sector of the
theory. The quark, lepton and Higgs fields are
\be
\begin{split}
Q^T=(u_L,d_L)~,~ u_R~,~d_R~,~\\L^T=(\nu_L,e_L)~,~e_R~,~ 
\nu_R~\mbox{ and }~H,
\end{split}
\ee
in an obvious notation. All of these fields have three species, and
we assume that each one forms a reducible representation ${\bf 1}_S\oplus{\bf 2}$.
The doublets carry capital indices $I$ and $J$, which run from $1$ to $2$,
and the singlets are denoted by
$Q_3,~u_{3R},~d_{3R},~L_3,~e_{3R},~\nu_{3R}$ and $~H_S$. Note that the subscript $3$ denotes the
singlet representation and not the third generation.
The most general renormalizable Yukawa interactions of this model are given by
\be
{\cal L}_Y = {\cal L}_{Y_D}+{\cal L}_{Y_U}
+{\cal L}_{Y_E}+{\cal L}_{Y_\nu},
\ee
where
\be
\begin{array}{lll}
{\cal L}_{Y_D} &=&
- Y_1^d \overline{ Q}_I H_S d_{IR} - Y_3^d \overline{ Q}_3 H_S d_{3R}  \\
&  &   -Y^{d}_{2}[~ \overline{ Q}_{I} \kappa_{IJ} H_1  d_{JR}
+\overline{ Q}_{I} \eta_{IJ} H_2  d_{JR}~]\\
&  & -Y^d_{4} \overline{ Q}_3 H_I  d_{IR} - Y^d_{5} \overline{ Q}_I H_I d_{3R} 
+~\mbox{h.c.} ,
\label{lagd}
\end{array}
\ee
\be
\begin{array}{lll}
{\cal L}_{Y_U} &=&
-Y^u_1 \overline{ Q}_{I}(i \sigma_2) H_S^* u_{IR} 
-Y^u_3\overline{ Q}_3(i \sigma_2) H_S^* u_{3R} \\
&  &   -Y^{u}_{2}[~ \overline{ Q}_{I} \kappa_{IJ} (i \sigma_2)H_1^*  u_{JR}
+\eta  \overline{ Q}_{I} \eta_{IJ}(i \sigma_2) H_2^*  u_{JR}~]\\
&  &
-Y^u_{4} \overline{ Q}_{3} (i \sigma_2)H_I^* u_{IR} 
-Y^u_{5}\overline{ Q}_I (i \sigma_2)H_I^*  u_{3R} +~\mbox{h.c.},
\label{lagu}
\end{array}
\ee
\be
\begin{array}{lll}
{\cal L}_{Y_E} &=& -Y^e_1\overline{ L}_I H_S e_{IR} 
-Y^e_3 \overline{ L}_3 H_S e_{3R} \\
&  &  - Y^{e}_{2}[~ \overline{ L}_{I}\kappa_{IJ}H_1  e_{JR}
+\overline{ L}_{I} \eta_{IJ} H_2  e_{JR}~]\\
 &  & -Y^e_{4}\overline{ L}_3 H_I e_{IR} 
-Y^e_{5} \overline{ L}_I H_I e_{3R} +~\mbox{h.c.},
\end{array}
\label{lage}
\ee
\be
\begin{array}{lcl}
{\cal L}_{Y_\nu} &=& -Y^{\nu}_1\overline{ L}_I (i \sigma_2)H_S^* \nu_{IR} 
-Y^\nu_3 \overline{ L}_3(i \sigma_2) H_S^* \nu_{3R} \\
&  &   -Y^{\nu}_{2}[~\overline{ L}_{I}\kappa_{IJ}(i \sigma_2)H_1^*  \nu_{JR}
+ \overline{ L}_{I} \eta_{IJ}(i \sigma_2) H_2^*  \nu_{JR}~]\\
 &  & -Y^\nu_{4}\overline{ L}_3(i \sigma_2) H_I^* \nu_{IR} 
-Y^\nu_{5} \overline{ L}_I (i \sigma_2)H_I^* \nu_{3R}+~\mbox{h.c.},
\label{lagnu}
\end{array}
\ee
and
\be
\kappa = \left( \begin{array}{cc}
0& 1\\ 1 & 0\\
\end{array}\right)~~\mbox{and}~~
\eta = \left( \begin{array}{cc}
1& 0\\ 0 & -1\\
\end{array}\right).
\label{kappa}
\ee Furthermore, we add to the Lagrangian the Majorana mass terms for
the right-handed neutrinos \be {\cal L}_{M} = -M_1 \nu_{IR}^T C
\nu_{IR} -M_3 \nu_{3R}^T C \nu_{3R}.
\label{majo}
\ee

Due to the presence of three Higgs fields, the Higgs potential
$V_H(H_S,H_D)$ is more complicated than that of the Standard
Model. This potential was analyzed by Pakvasa and Sugawara~\cite{pakvasa1} who
found that in addition to the $S_{3}$ symmetry, it has a permutational
symmetry
$S_{2}$: $H_{1}\leftrightarrow H_{2}$, which is not a subgroup of the
flavour group $S_{3}$, and an Abelian discrete symmetry that we will use
for selection rules of the Yukawa couplings in the leptonic sector. In
this communication, we will assume that the vacuum respects the
accidental $S_{2}$ symmetry of the Higgs potential and that
\be
\langle H_{1} \rangle = \langle H_{2} \rangle.
\ee

With these assumptions, the Yukawa interactions, eqs. (\ref{lagd})-(\ref{lagnu}) yield mass matrices,
for all fermions in the theory, of the general form
\be
{\bf M} = \left( \begin{array}{ccc}
\mu_{1}+\mu_{2} & \mu_{2} & \mu_{5} 
\\  \mu_{2} & \mu_{1}-\mu_{2} &\mu_{5}
  \\ \mu_{4} & \mu_{4}&  \mu_{3}
\end{array}\right).
\label{general-m}
\ee
The Majorana mass for the left handed neutrinos $\nu_{L}$ is generated
by the see-saw mechanism. The corresponding mass matrix is
given by
\be
{\bf M_{\nu}} = {\bf M_{\nu_D}}\tilde{{\bf M}}^{-1}({\bf M_{\nu_D}})^T,
\label{seesaw}
\ee
where $\tilde{{\bf M}}=\mbox{diag}(M_1,M_1,M_3)$.
\\
In principle, all entries in the mass matrices can be complex since
there is no restriction coming from the flavour symmetry $S_{3}$.
The mass matrices are diagonalized by bi-unitary transformations as
\be
\begin{array}{rcl}
U_{d(u,e)L}^{\dag}{\bf M}_{d(u,e)}U_{d(u,e)R} 
&=&\mbox{diag} (m_{d(u,e)}, m_{s(c,\mu)},m_{b(t,\tau)}),
\\ 
\\
U_{\nu}^{T}{\bf M_\nu}U_{\nu} &=&
\mbox{diag} (m_{\nu_1},m_{\nu_2},m_{\nu_3}).
\end{array}
\label{unu}
\ee
The entries in the diagonal matrices may be complex, so the physical
masses are their absolute values.

The mixing matrices are, by definition,
\be
\begin{array}{ll}
V_{CKM} = U_{uL}^{\dag} U_{dL},& V_{PMNS} = U_{eL}^{\dag} U_{\nu} K.
\label{ckm1}
\end{array}
\ee
where $K$ is the diagonal matrix of the Majorana phase factors.
\section{The mass matrices in the leptonic sector and $Z_{2}$
  symmetry}
A further reduction of the number of parameters in the leptonic sector
may be achieved by means of an Abelian $Z_{2}$ symmetry. A possible set
of charge assignments of $Z_{2}$, compatible with the experimental
data on masses and mixings in the leptonic sector is given in Table I.
\begin{center}
\begin{tabular}{|c|c|}
\hline
 $-$ &  $+$
\\ \hline

$H_S, ~\nu_{3R}$ & $H_I, ~L_3, ~L_I, ~e_{3R},~ e_{IR},~\nu_{IR}$
\\ \hline
\end{tabular}
\end{center}
\vspace{-0.3cm}
\begin{center}
{\footnotesize {\bf Table I}. $Z_2$ assignment in the leptonic sector.}
\end{center}
These $Z_2$ assignments forbid the following Yukawa couplings
\be
 Y^e_{1} = Y^e_{3}= Y^{\nu}_{1}= Y^{\nu}_{5}=0.
\label{zeros}
\ee
Therefore, the corresponding entries in the mass matrices vanish, {\it
  i.e.}, $\mu_{1}^{e}=\mu_{3}^{e}=0$ and $\mu_{1}^{\nu}=\mu_{5}^{\nu}=0$.
\begin{center}{\it The mass matrix of the charged leptons}\end{center}
The mass matrix of the charged leptons takes the form
\be
M_{e} = m_{\tau}\left( \begin{array}{ccc}
\tilde{\mu}_{2} & \tilde{\mu}_{2} & \tilde{\mu}_{5} 
\\  \tilde{\mu}_{2} &-\tilde{\mu}_{2} &\tilde{\mu}_{5}
  \\ \tilde{\mu}_{4} & \tilde{\mu}_{4}& 0
\end{array}\right).
\label{charged-leptons-m}
\ee
The unitary matrix $U_{eL}$ that enters in the definition of the
mixing matrix, $V_{PMNS}$, is calculated from
\be
U_{eL}^{\dag}M_{e}M_{e}^{\dag}U_{eL}=\mbox{diag}(m_{e}^{2},m_{\mu}^{2},m_{\tau}^{2}),
\ee
where $m_{e}$, $m_{\mu}$ and $m_{\tau}$ are the masses of the charged
leptons, and
\begin{equation}
M_{e}M_{e}^{\dag}=m_{\tau}^2 \left( \begin {array}{ccc} 2|\tilde{\mu}_{2}|^2+|\tilde{\mu}_{5}|^{2}&|\tilde{\mu}_{5}|^{2}&2|\tilde{\mu}_{2}||\tilde{\mu}_{4}|e^{-i\delta_{e}}\\\noalign{\medskip}|\tilde{\mu}_{5}|^{2}&2|\tilde{\mu}_{2}|^{2}+|\tilde{\mu}_{5}|^{2}&0\\\noalign{\medskip}2|\tilde{\mu}_{2}||\tilde{\mu}_{4}|e^{i\delta_{e}}&0&2\,|\tilde{\mu}_{4}|^{2}\end {array} \right).
\label{mmdag}
\end{equation}
Notice that this matrix has only one non-ignorable phase factor.
The parameters $|\tilde{\mu}_{2}|$, $|\tilde{\mu}_{4}|$ and
$|\tilde{\mu}_{5}|$ may readily be expressed in terms of the charged
lepton masses. From the invariants of $M_{e}M_{e}^{\dag}$, we get the
set of equations
\be
 Tr(M_{e}M_{e}^{\dag})=m_{e}^{2}+m_{\mu}^{2}+m_{\tau}^{2}
=m_{\tau}^{2}\left[4|\tilde{\mu}_{2}|^{2}+2\left(|\tilde{\mu}_{4}|^{2}+|\tilde{\mu}_{5}|^{2}\right)
  \right],
\label{trace}
\ee

\be
\begin{array}{lcl}
\chi(M_{e}M_{e}^{\dag})&=&m_{\tau}^{2}(m_{e}^{2}+m_{\mu}^{2})+m_{e}^{2}m_{\mu}^{2}\\
&=&4m_{\tau}^{4}\left[
|\tilde{\mu}_{2}|^{4}+|\tilde{\mu}_{2}|^{2}\left(|\tilde{\mu}_{4}|^{2}+|\tilde{\mu}_{5}|^{2}\right)+|\tilde{\mu}_{4}|^{2}|\tilde{\mu}_{5}|^{2}
\right],
\end{array}
\label{chis}
\ee
\be
\begin{array}{l}
det(M_{e}M_{e}^{\dag})=m_{e}^{2}m_{\mu}^{2}m_{\tau}^{2}
=4m_{\tau}^{6}|\tilde{\mu}_{2}|^{2}|\tilde{\mu}_{4}|^{2}|\tilde{\mu}_{5}|^{2},
\end{array}
\label{determinant}
\ee
where {\small
$\chi(M_{e}M_{e}^{\dag})=\frac{1}{2}\left[(Tr(M_{e}M_{e}^{\dag}))^{2}-Tr(M_{e}M_{e}^{\dag})^2
\right]$}.

Solving these equations for $|\tilde{\mu}_{2}|^2$, $|\tilde{\mu}_{4}|^2$ and
$|\tilde{\mu}_{5}|^2$, we obtain
\be
|\tilde{\mu}_{2}|^2=\frac{1}{2}\frac{m_{e}^{2}+m_{\mu}^{2}}{m_{\tau}^{2}}-\frac{m_{e}^{2}m_{\mu}^{2}}{m_{\tau}^{2}(m_{e}^{2}+m_{\mu}^{2})}+\beta.
\label{mu2beta}
\ee
In this expression, $\beta$ is the smallest solution of the equation
\begin{equation}
\begin{array}{l}
\beta^3-\frac{1}{2}(1-2y+6\frac{z}{y})\beta^2-\frac{1}{4}(y-y^{2}-4\frac{z}{y}+7z-12\frac{z^{2}}{y^{2}})\beta-
\\ \\
\frac{1}{8}yz-\frac{1}{2}\frac{z^{2}}{y^{2}}+\frac{3}{4}\frac{z^{2}}{y}-\frac{z^{3}}{y^{3}}=0,
\label{cubic4beta}
\end{array}
\end{equation}
where $y=(m_{e}^{2}+m_{\mu}^2)/m_{\tau}^{2}$ and $z=m_{\mu}^2m_{e}^{2}/m_{\tau}^{4}$.\\
A good, order of magnitude, estimation for $\beta$ is obtained from
(\ref{cubic4beta})
\be
\beta \approx
-\frac{m_{\mu}^2m_{e}^{2}}{2m_{\tau}^{2}(m_{\tau}^{2}-(m_{\mu}^2+m_{e}^{2}))}.
\ee
The parameters $|\tilde{\mu}_{4}|^2$ and $|\tilde{\mu}_{5}|^2$ are,
then, readily expressed in terms of $|\tilde{\mu}_{2}|^2$,
\be
\begin{split}
\begin{array}{l}
|\tilde{\mu}_{4,5}|^2=
\frac{1}{4}\left(1-\frac{m_{\mu}^2+m_{e}^{2}}{m_{\tau}^{2}}+4\frac{m_{e}^{2}m_{\mu}^{2}}{m_{\tau}^{2}(m_{e}^{2}+m_{\mu}^{2})}-4\beta
\right)\\ \\
\pm\frac{1}{4}\left(\sqrt{({\scriptsize 1-\frac{m_{\mu}^2+m_{e}^{2}}{m_{\tau}^{2}}+4\frac{m_{e}^{2}m_{\mu}^{2}}{m_{\tau}^{2}(m_{e}^{2}+m_{\mu}^{2})}-4\beta})^2-\frac{m_{\mu}^2m_{e}^{2}}{m_{\tau}^{4}}\frac{1}{|\tilde{\mu}_{2}|^2}}\right).
\end{array}
\end{split}
\ee
Once $M_{e}M_{e}^{\dag}$ has been reparametrized in terms of the
charged lepton masses, it is straightforward to compute $U_{eL}$ also
as a function of the lepton masses. 
Here, in order to avoid a clumsy notation, we will write the result to
order $\left(m_{\mu}m_{e}/m_{\tau}^{2}\right)^{2}$ and $x^{4}=\left(m_{e}/m_{\mu}\right)^4$
\be
M_{e}\approx m_{\tau} \left( 
\begin{array}{ccc}
\frac{1}{\sqrt{2}}\frac{\tilde{m}_{\mu}}{\sqrt{1+x^2}} & \frac{1}{\sqrt{2}}\frac{\tilde{m}_{\mu}}{\sqrt{1+x^2}} & \frac{1}{\sqrt{2}} \sqrt{\frac{1+x^2-\tilde{m}_{\mu}^2}{1+x^2}} \\ \\
 \frac{1}{\sqrt{2}}\frac{\tilde{m}_{\mu}}{\sqrt{1+x^2}} &-\frac{1}{\sqrt{2}}\frac{\tilde{m}_{\mu}}{\sqrt{1+x^2}}  & \frac{1}{\sqrt{2}} \sqrt{\frac{1+x^2-\tilde{m}_{\mu}^2}{1+x^2}} \\ \\
\frac{\tilde{m}_{e}(1+x^2)}{\sqrt{1+x^2-\tilde{m}_{\mu}^2}}e^{i\delta_{e}} & \frac{\tilde{m}_{e}(1+x^2)}{\sqrt{1+x^2-\tilde{m}_{\mu}^2}}e^{i\delta_{e}} & 0
\end{array}
\right).
\label{emass}
\ee
This approximation is numerically exact up to order $10^{-9}$ in units
of the $\tau$ mass.

The unitary matrix $U_{eL}$ that diagonalizes $M_{e}M_{e}^{\dagger}$ and
enters in the definition of the neutrino mixing matrix $V_{PMNS}$,
eq. (\ref{ckm1}), is
\be
\ba{l}
U_{eL}\approx \left(\ba{ccc} 
1& 0 & 0 \\
0 & 1 & 0 \\
0 & 0 & e^{i\delta_{e}}
\ea\right) \left(
\ba{ccc}
O_{11}& -O_{12}& O_{13} \\
-O_{21}& O_{22}& O_{23} \\
-O_{31}& -O_{32}& O_{33} 
\ea
\right)~,
\ea
\label{unitary-leptons}
\ee
where
\be
\begin{array}{l}
\left(
\ba{ccc}
O_{11}& -O_{12}& O_{13} \\
-O_{21}& O_{22}& O_{23} \\
-O_{31}& -O_{32}& O_{33} 
\ea
\right)=\\\\
\left(
\ba{ccc}
\frac{1}{\sqrt{2}}x
\frac{(
1+2\tilde{m}_{\mu}^2+4x^2+\tilde{m}_{\mu}^4+2\tilde{m}_{e}^2
)}{\sqrt{1+\tilde{m}_{\mu}^2+5x^2-\tilde{m}_{\mu}^4-\tilde{m}_{\mu}^6+\tilde{m}_{e}^2+12x^4}}&
-\frac{1}{\sqrt{2}}\frac{(1-2\tilde{m}_{\mu}^2+\tilde{m}_{\mu}^4-2\tilde{m}_{e}^2)}{\sqrt{1-4\tilde{m}_{\mu}^2+x^2+6\tilde{m}_{\mu}^4-4\tilde{m}_{\mu}^6-5\tilde{m}_{e}^2}}
& \frac{1}{\sqrt{2}} \\ \\
-\frac{1}{\sqrt{2}}x
\frac{(
1+4x^2-\tilde{m}_{\mu}^4-2\tilde{m}_{e}^2
)}{\sqrt{1+\tilde{m}_{\mu}^2+5x^2-\tilde{m}_{\mu}^4-\tilde{m}_{\mu}^6+\tilde{m}_{e}^2+12x^4}}
&
\frac{1}{\sqrt{2}}\frac{(1-2\tilde{m}_{\mu}^2+\tilde{m}_{\mu}^4)}{\sqrt{1-4\tilde{m}_{\mu}^2+x^2+6\tilde{m}_{\mu}^4-4\tilde{m}_{\mu}^6-5\tilde{m}_{e}^2}}
& \frac{1}{\sqrt{2}} \\ \\
-\frac{\sqrt{1+2x^2-\tilde{m}_{\mu}^2-\tilde{m}_{e}^2}(1+\tilde{m}_{\mu}^2+x^2-2\tilde{m}_{e}^2)}{\sqrt{1+\tilde{m}_{\mu}^2+5x^2-\tilde{m}_{\mu}^4-\tilde{m}_{\mu}^6+\tilde{m}_{e}^2+12x^4}} & -x\frac{(1+x^2-\tilde{m}_{\mu}^2-2\tilde{m}_{e}^2)\sqrt{1+2x^2-\tilde{m}_{\mu}^2-\tilde{m}_{e}^2}}{\sqrt{1-4\tilde{m}_{\mu}^2+x^2+6\tilde{m}_{\mu}^4-4\tilde{m}_{\mu}^6-5\tilde{m}_{e}^2}} &\tilde{m}_{e}\tilde{m}_{\mu}\frac{\sqrt{1+x^2}}{\sqrt{1+x^2-\tilde{m}_{\mu}^2}}
\ea
\right)~,
\label{unitary-leptons-2}
\end{array}
\ee
and where $\tilde{m_{\mu}}=m_{\mu}/m_{\tau}$,
$\tilde{m_{e}}=m_{e}/m_{\tau}$ and $x=m_{e}/m_{\mu}$.
\vspace{-0.1cm}
\begin{center}{\it The mass matrix of the neutrinos}\end{center}
According with the $Z_{2}$ selection rule eq. (\ref{zeros}), the mass
matrix of the Dirac neutrino takes the form

\be
{\bf M_{\nu_D}} = \left( \begin{array}{ccc}
\mu^{\nu}_{2} & \mu^{\nu}_{2} & 0
\\  \mu^{\nu}_{2} & -\mu^{\nu}_{2} &0
  \\ \mu^{\nu}_{4} & \mu^{\nu}_{4}&  \mu^{\nu}_{3}
\end{array}\right).
\label{neutrinod-m}
\ee
\\
Then, the mass matrix for the left-handed Majorana neutrinos, obtained
from the see-saw mechanism, is

\be
{\bf M_{\nu}} = {\bf M_{\nu_D}}\tilde{{\bf M}}^{-1} 
({\bf M_{\nu_D}})^T=
\left( \begin{array}{ccc}
2 (\rho^{\nu}_{2})^2 & 0 & 
2 \rho^{\nu}_2 \rho^{\nu}_{4}
\\ 0 & 2 (\rho^{\nu}_{2})^2 & 0
  \\ 2 \rho^{\nu}_2 \rho^{\nu}_{4} & 0  &  
2 (\rho^{\nu}_{4})^2 +
(\rho^{\nu}_3)^2
\end{array}\right),
\label{m-nu}
\ee
where $\rho_2^\nu =(\mu^{\nu}_2)/M_{1}^{1/2}$,  $\rho_4^\nu
=(\mu^{\nu}_4)/M_{1}^{1/2}$ and $\rho_3^\nu
=(\mu^{\nu}_3)/M_{3}^{1/2}$; $M_{1}$ and $M_{3}$ are the masses of
the right handed neutrinos appearing in (\ref{majo}).

The non-Hermitian, complex, symmetric neutrino mass matrix $M_{\nu}$ may be brought
to a diagonal form by a bi-unitary transformation, as
\be
U_{\nu}^{T}M_{\nu}U_{\nu}=\mbox{diag}\left(|m_{\nu_{1}}|e^{i\phi_{1}},|m_{\nu_{2}}|e^{i\phi_{2}},|m_{\nu_{3}}|e^{i\phi_{\nu}}\right),
\label{diagneutrino}
\ee
where $U_{\nu}$ is the matrix that diagonalizes the matrix
$M_{\nu}^{\dagger}M_{\nu}$.\\
In order to compute $U_{\nu}$, we notice that $M_{\nu}^{\dagger}M_{\nu}$ has the same
texture zeroes as $M_{\nu}$
\be
M_{\nu}^{\dagger}M_{\nu}=\left(
\ba{ccc}
|A|^2+ |B|^2 & 0 & A^{\star}B+B^{\star}D \\
0 & |A|^2 & 0 \\
AB^{\star}+BD^{\star} & 0 &  |B|^2+|D|^2
\ea\right),
\ee
where $A=2 (\rho^{\nu}_{2})^2$, $B=2 \rho^{\nu}_2 \rho^{\nu}_{4}$, and $D=2 (\rho^{\nu}_{4})^2 +
(\rho^{\nu}_3)^2$.\\
Furthermore, notice that the entries in the upper right corner and lower left
corner are complex conjugates of each other, all other entries are
real. Therefore, the matrix $U_{\nu L}$ that diagonalizes
$M_{\nu}^{\dagger}M_{\nu}$, takes the form
\be
U_{\nu}=\left(\ba{ccc} 
1& 0 & 0 \\
0 & 1 & 0 \\
0 & 0 & e^{i\delta_{\nu}} 
\ea\right)\left(
\begin{array}{ccc}
\cos \eta & \sin \eta & 0 \\
0 & 0 & 1 \\
-\sin \eta  & \cos \eta &0
\end{array}
\right).
\label{ununew}
\ee
If we require that the defining equation (\ref{diagneutrino}) be
satisfied as an identity, we get the following set of equations:
\be
\ba{l}
2 (\rho^{\nu}_{2})^2=m_{\nu_{3}},\\
\\
2 (\rho^{\nu}_{2})^2=m_{\nu_{1}}\cos^2 \eta + m_{\nu_{2}}\sin^2 \eta, \\
\\
2 \rho^{\nu}_2 \rho^{\nu}_{4}=\sin \eta \cos \eta (m_{\nu_{2}}-m_{\nu_{1}})e^{-i\delta_{\nu}},\\
\\
2 (\rho^{\nu}_{4})^2 +
(\rho^{\nu}_3)^2=(m_{\nu_{1}}\sin^2 \eta + m_{\nu_{2}}\cos^2 \eta)e^{-2i\delta_{\nu}}.
\ea
\ee
Solving these equations for $\sin \eta$ and $\cos \eta$, we find
\be
\ba{l}
\sin^2\eta=\frac{m_{\nu_{3}}-m_{\nu_{1}}}{m_{\nu_{2}}-m_{\nu_{1}}},\\
\\
\cos^2\eta=\frac{m_{\nu_{2}}-m_{\nu_{3}}}{m_{\nu_{2}}-m_{\nu_{1}}}.
\ea
\label{cosysin}
\ee
The unitarity of $U_{\nu}$ constrains $\sin \eta$ to be real and thus 
$|\sin \eta|\leq 1$, this condition fixes the phases $\phi_{1}$ and
$\phi_{2}$ as
\be
|m_{\nu_{1}}|\sin \phi_{1}=|m_{\nu_{2}}|\sin \phi_{2}=|m_{\nu_{3}}|\sin \phi_{\nu}.
\label{phase-condition}
\ee
The real phase $\delta_{\nu}$ appearing in eq. (\ref{ununew}) is not
constrained by the unitarity of $U_{\nu}$.

Substitution of the expressions (\ref{cosysin}) for $\sin \eta$ and
$\cos \eta$ in (\ref{ununew}) allows us to write the unitary matrix
$U_{\nu}$ as
\be
U_{\nu}=
\left(\ba{ccc} 
1& 0 & 0 \\
0 & 1 & 0 \\
0 & 0 & e^{i\delta_{\nu}} 
\ea\right)\left(
\begin{array}{ccc}
\sqrt{\displaystyle{\frac{m_{\nu_{2}}-m_{\nu_{3}}}
{m_{\nu_{2}}-m_{\nu_{1}}}}}& 
\sqrt{
\displaystyle{\frac{m_{\nu_{3}}-m_{\nu_{1}}}{m_{\nu_{2}}-m_{\nu_{1}}}}} & 0\\
0&0&1\\
-\sqrt{
\displaystyle{\frac{m_{\nu_{3}}-m_{\nu_{1}}}{m_{\nu_{2}}-m_{\nu_{1}}}}}  &\sqrt{\displaystyle{\frac{m_{\nu_{2}}-m_{\nu_{3}}}
{m_{\nu_{2}}-m_{\nu_{1}}}}}&0
\end{array}
\right).
\label{unu-final}
\ee
\\
Now, the mass matrix of the Majorana neutrinos, $M_{\nu}$, may be
written in terms of the neutrino masses; from (\ref{diagneutrino})
and (\ref{unu-final}), we get
\be
M_{\nu} = 
\left( \begin{array}{ccc}
m_{\nu_{3}} & 0 & \sqrt{(m_{\nu_{3}}-m_{\nu_{1}})(m_{\nu_{2}}-m_{\nu_{3}})}e^{-i\delta_{\nu}}
\\ 
0 &m_{\nu_{3}}  & 0
\\
\sqrt{(m_{\nu_{3}}-m_{\nu_{1}})(m_{\nu_{2}}-m_{\nu_{3}})} e^{-i\delta_{\nu}} & 0  & (m_{\nu_{1}}+m_{\nu_{2}}-m_{\nu_{3}})e^{-2i\delta_{\nu}}
\end{array}\right).
\label{m-nu2}
\ee
The only free parameters in these matrices, other than the neutrino
masses, are the phase $\phi_{\nu}$, implicit in $m_{\nu_{1}}$,
$m_{\nu_{2}}$ and $m_{\nu_{3}}$, and the Dirac phase $\delta_{\nu}$.

\begin{center}
{\it The neutrino mixing matrix}
\end{center}
The neutrino mixing matrix $V_{PMNS}$, is the product
$U_{eL}U_{\nu}^{\dagger}K$, where $K$ is the diagonal matrix of the
Majorana phase factors, defined by
\be
diag(m_{\nu_{1}},m_{\nu_{2}},m_{\nu_{3}})=K^{\dagger}diag(|m_{\nu_{1}}|,|m_{\nu_{2}}|,|m_{\nu_{3}}|)K^{\dagger}.
\ee
Except for an overall phase factor $e^{i\phi_{1}}$, which can be
ignored, $K$ is 
\be
K=diag(1,e^{i\alpha},e^{i\beta}),
\ee
where $\alpha=1/2(\phi_{1}-\phi_{2})$ and
$\beta=1/2(\phi_{1}-\phi_{\nu})$ are the Majorana phases. The neutrino mixing matrix $V_{PMNS}$, in the standard form advocated
by the $PDG$~\cite{PDG}, is obtained by taking the product
$U_{eL}^{\dag}U_{\nu}K$ and making an appropriate transformation of
phases. Writing the resulting expression to the same approximation as
in eq. (\ref{emass}), we get
\be
\begin{split}
V_{PMNS}\approx
\left(
\ba{ccc}
O_{11}\cos \eta + O_{31}\sin \eta  & O_{11}\sin
\eta-O_{31} \cos \eta  & -O_{21} e^{-i\delta} \\ \\
-O_{12}\cos \eta + O_{32}\sin \eta e^{i\delta} & -O_{12}\sin
\eta-O_{32}\cos \eta e^{i\delta} & O_{22} \\ \\
O_{13}\cos \eta - O_{33}\sin \eta e^{i\delta} & O_{13}\sin
\eta+O_{33}\cos \eta e^{i\delta} & O_{23} 
\ea
\right)
 \times K,
\end{split}
\ee
where $\cos \eta$ and $\sin \eta$ are given in eq (\ref{cosysin}),
$O_{ij}$ are given in eq (\ref{unitary-leptons-2}) and $\delta=\delta_{\nu}-\delta_{e}$.

A comparison of this expression with the standard parametrization
allowed us to derive expressions for the mixing angles in terms of the
charged lepton and neutrino masses 
\be
\ba{llll}
\sin \theta_{13}=-O_{21}
, &
\sin \theta_{23}= \frac{O_{22}}{\sqrt{O_{22}^2+O_{23}^2}} & \mbox{and}
& \tan \theta_{12}=\frac{O_{11}\sin \eta-O_{31}\cos \eta }{O_{31}\sin \eta+O_{11}\cos \eta}.
\ea
\ee
Keeping only terms of order $(m_{e}^2/m_{\mu}^2)$ and
$(m_{\mu}/m_{\tau})^4$, we get
\be
\ba{lr}
\sin \theta_{13}\approx -\frac{1}{\sqrt{2}}x
\frac{(
1+4x^2-\tilde{m}_{\mu}^4)}{\sqrt{1+\tilde{m}_{\mu}^2+5x^2-\tilde{m}_{\mu}^4}}
, &
\sin \theta_{23}\approx  \frac{1}{\sqrt{2}}\frac{1-2\tilde{m}_{\mu}^2+\tilde{m}_{\mu}^4}{\sqrt{1-4\tilde{m}_{\mu}^2+x^2+6\tilde{m}_{\mu}^4}}
\ea
\ee
and
\be
\begin{array}{l}
\tan \theta_{12}= -\sqrt{\frac{\displaystyle{m_{\nu_{2}}-m_{\nu_{3}}}}{
\displaystyle{m_{\nu_{3}}-m_{\nu_{1}}}}}\times \left(
\frac{\sqrt{1+2x^2-\tilde{m}_{\mu}^2}(1+\tilde{m}_{\mu}^2+x^2)-
\frac{1}{\sqrt{2}}x(
1+2\tilde{m}_{\mu}^2+4x^2)
\sqrt{\frac{
\displaystyle{m_{\nu_{3}}-m_{\nu_{1}}}}{\displaystyle{m_{\nu_{2}}-m_{\nu_{3}}}}}}{\sqrt{1+2x^2-\tilde{m}_{\mu}^2}(1+\tilde{m}_{\mu}^2+x^2)+
\frac{1}{\sqrt{2}}x(
1+2\tilde{m}_{\mu}^2+4x^2)
\sqrt{\frac{\displaystyle{m_{\nu_{2}}-m_{\nu_{3}}}}{
\displaystyle{m_{\nu_{3}}-m_{\nu_{1}}}}}}
\right).
\end{array}
\label{tan}
\ee
The dependence of $\tan \theta_{12}$ on the phase $\phi_{\nu}$ and the
physical masses of the neutrinos enters through the ratio of the
neutrino mass differences under the square root sign, it can be made
explicit with the help of the unitarity constraint on $U_{\nu}$, 
eq. (\ref{phase-condition}),
\be
\frac{\displaystyle{m_{\nu_{2}}-m_{\nu_{3}}}}{
\displaystyle{m_{\nu_{3}}-m_{\nu_{1}}}}=
\frac{(|m_{\nu_{2}}|^2-|m_{\nu_{3}}|^{2}\sin^{2}\phi_{\nu})^{1/2}-|m_{\nu_{3}}|
  |\cos
    \phi_{\nu}|}
{(|m_{\nu_{1}}|^{2}-|m_{\nu_{3}}|^{2}\sin^{2}\phi_{\nu})^{1/2}+|m_{\nu_{3}}|
  |\cos
    \phi_{\nu}|}.
\label{tansq}
\ee
Similarly, the Majorana phases are given by
\be
\begin{array}{l}
\sin 2\alpha=\sin(\phi_{1}-\phi_{2})=
\frac{|m_{\nu_{3}}|\sin\phi_{\nu}}{|m_{\nu_{1}}||m_{\nu_{2}}|}\times
\left(\sqrt{|m_{\nu_{2}}|^2-|m_{\nu_{3}}|^{2}\sin^{2}\phi_{\nu}}+\sqrt{|m_{\nu_{1}}|^{2}-|m_{\nu_{3}}|^{2}\sin^{2}\phi_{\nu}}\right),
\\
\\
\sin 2\beta=\sin(\phi_{1}-\phi_{\nu})=
 \frac{\sin\phi_{\nu}}{|m_{\nu_{1}}|}\left(|m_{\nu_{3}}|\sqrt{1-\sin^{2}\phi_{\nu}}+\sqrt{|m_{\nu_{1}}|^{2}-|m_{\nu_{3}}|^{2}\sin^{2}\phi_{\nu}}\right).
\end{array}
\ee
A more complete and detailed discussion of the Majorana phases in the
neutrino mixing matrix $V_{PMNS}$ obtained in our model is given by 
J. Kubo~\cite{kubo-u}.

In the present model, $\sin^{2} \theta_{13}$ and $\sin^{2} \theta_{23}$ are
determined by the masses of the charged leptons in very good
agreement with the experimental values~\cite{Maltoni:2004ei,schwetz,fogli1},
\be
\begin{array}{ll}
(\sin^{2}\theta_{13})^{th}=1.1\times 10^{-5}, &(\sin^2
  \theta_{13})^{exp} \leq 0.046, \nonumber
\end{array}
\ee
and
\be
\begin{array}{ll}
(\sin^{2}\theta_{23})^{th}=0.499, &(\sin^2
  \theta_{23})^{exp}=0.5^{+0.06}_{-0.05}.\nonumber
\end{array}
\ee
thus, the experimental restriction $|\Delta
m^2_{12}|<|\Delta m^2_{13}|$ implies an inverted neutrino mass
spectrum, $|m_{\nu_{3}}|<|m_{\nu_{1}}|<|m_{\nu_{2}}|$~\cite{kubo1}.

As seen from eqs. (\ref{tan}) and (\ref{tansq}), the solar mixing angle is
sensitive to the neutrino mass differences and the phase $\phi_{\nu}$
but is only very weakly sensitive to the charged lepton
masses. Writing the neutrino mass differences
$m_{\nu_{i}}-m_{\nu_{j}}$ in terms of the differences of the mass
squared and one of the neutrino masses, say $|m_{\nu_{2}}|$, from
our previous expressions (\ref{tan}) and (\ref{tansq}), we obtain
\be
\begin{split}
\begin{array}{lll}
\frac{m_{\nu_{2}}^{2}}{\Delta
  m^2_{13}}&=&\frac{1+2t_{12}^{2}+t^{4}_{12}-rt^{4}_{12}}{4t^{2}_{12}(1+t^{2}_{12})(1+t^{2}_{12}-r
  t^{2}_{12})cos^{2}\phi_{\nu}} - \tan^{2} \phi_{\nu} +
  O\left(\frac{m_{e}^{2}}{m_{\mu}^{2}}\right)\\
\\
&\approx&
  \frac{1}{\sin^{2}2\theta_{12}\cos^{2}\phi_{\nu}}-\tan^{2} \phi_{\nu}
  ~\mbox{   for   }~ r << 1,
\end{array}
\end{split}
\ee
where $t_{12}=\tan \theta_{12}$ and $r=\Delta
  m^2_{21}/\Delta
  m^2_{13}$.

The mass $|m_{\nu_{2}}|$ assumes its minimal value when $\sin
\phi_{\nu}$ vanishes, then
\be
|m_{\nu_{2}}|\approx \frac{\sqrt{\Delta
  m^2_{13}}}{\sin 2\theta_{12}}.
\ee
Hence, we find
\be
\begin{split}
|m_{\nu_{2}}|\approx0.0507eV,\\ |m_{\nu_{1}}|\approx 0.0499eV,\\
|m_{\nu_{3}}|\approx 0.0193eV,
\end{split}
\ee
where we used the values $\Delta m^{2}_{13}=2.2^{+0.37}_{-0.27}\times
10^{-3}eV^{2}$ and $\sin^{2}\theta_{12}=0.31^{+0.02}_{-0.03}$ taken
from M. Maltoni et al.~\cite{Maltoni:2004ei}, T. Schwetz~\cite{schwetz} and
G. L. Fogli et al.~\cite{fogli1}.

With those values for the neutrino masses we compute the effective
electron neutrino mass $m_{\beta}$
\begin{equation}
 \label{mb} m_\beta = \left[\sum_i|U_{ei}|^2m^2_{\nu_{i}}\right]^\frac{1}{2}=0.0502eV ,
\end{equation}
well below the upper bound $m_{\beta}<1.8eV$ coming from the tritium
$\beta$-decay experiments~\cite{eitel,fogli1,serra}.

\section{Conclusions}
By introducing three Higgs fields that are $SU(2)_{L}$ doublets in the
theory, we extended the concept of flavour and generations to the
Higgs sector and formulated a Minimal $S_{3}$-Invariant Extension of
the Standard Model~\cite{kubo1}. A well defined structure of the Yukawa
couplings is obtained, which permits the calculation of mass and mixing
matrices for quarks and leptons in a unified way. A further reduction
of redundant parameters is achieved in the leptonic sector by
introducing a $Z_{2}$ symmetry. The flavour symmetry group $Z_{2}
\times S_{3}$ relates the mass spectrum and mixings. This allowed us
to compute the neutrino mixing matrix explicitly in terms of the
masses of the charged leptons and neutrinos. In this model, the magnitudes
of the three mixing angles are determined by the interplay of the
flavour $S_{3}\times Z_{2}$ symmetry, the see-saw mechanism and the
lepton mass hierarchy. We also found that $V_{PMNS}$ has three CP
violating phases, one Dirac phase
$\delta=\delta_{\nu}-\delta_{e}$ and two 
Majorana phases, $\alpha$ and
$\beta$, that are functions of the neutrino masses, and another phase
$\phi_{\nu}$ which is independent of the Dirac phase. The numerical
values of the reactor, $\theta_{13}$, and the atmospheric,
$\theta_{23}$, mixing angles are determined by the masses of the
charged leptons only, in very good agreement with the experiment. The
solar mixing angle $\theta_{12}$ is almost insensitive to the values
of the masses of the charged leptons, but its experimental value
allowed us to fix the scale and origin of the neutrino mass spectrum,
which has an inverted hierarchy, with the values $|m_{\nu_{2}}|=0.0507eV$,
$|m_{\nu_{1}}|=0.0499eV$ and $|m_{\nu_{3}}|=0.0193eV$.
\section{Acknowledgements}
We are indebted to Prof. C. S. Lam for calling our attention to some
typographical errors in the previous version of this paper. This work
was partially supported by CONACYT M\'exico under contract No 40162-F
and by DGAPA-UNAM under contract PAPIIT-IN116202-3.

\end{document}